\documentclass[twocolumn,nofootinbib]{revtex4}

\begin{document}

\title{Can the cosmological dark sector be modeled by a single scalar field?}

\author{Saulo Carneiro}

\affiliation{Instituto de F\'{\i}sica Gleb Wataghin - UNICAMP, 13083-970, Campinas, S\~ao Paulo, Brazil\\ Instituto de F\'{\i}sica, Universidade Federal da Bahia, 40210-340, Salvador, Bahia, Brazil}

\begin{abstract}
In a previous paper it was shown that a minimally coupled scalar field of mass $M \sim H_0$ can describe both components of the dark sector in a unified way. In the solution found, the dark energy component decays linearly with the Hubble parameter, with a homogeneous creation of dark matter. In the present note we show that a $\Lambda$CDM dark sector can also be modeled by such a single field. More generally, we show that the system of Klein-Gordon and Einstein equations admits a uniparametric family of solutions that is equivalent to a non-adiabatic (with zero sound speed) generalised Chaplygin gas.
\end{abstract}

\maketitle

In spite of the increasing number of  precise observations pointing to the presence of two dark components in the cosmic fluid, our theoretical knowledge about them has not increased in the same rate \cite{teoricos}. The simplest description of the dark sector is provided by the $\Lambda$CDM model, with a cosmological constant and cold dark matter. Aside the absence of explanation for the energy scale of the former, this model has also presented some observational tensions \cite{tensions} that have lead to looking for alternatives \cite{alternatives}. On the other hand, since both dark components have no baryonic origin, the hypothesis of a common origin for them has been explored in unified schemes like, for example, generalised Chaplygin gas models \cite{gCg} and unified descriptions based on scalar fields \cite{scalar}. Unified dark fluids have, however, the undesired property of presenting a non-null sound velocity, which leads to instabilities and oscillations in the power spectrum of density fluctuations \cite{waga1}. This bad behaviour can be avoided if we properly identify, in the unified fluid, a strictly homogeneous component responsible for the cosmic acceleration - which we can name dark energy - and a pressureless component to be identified with clustering dark matter observed in galaxies and clusters \cite{non-adiabatic,borges}. In this process we derive an interacting dark sector \cite{iModels}, with a flux of energy between the two defined components\footnote{This is an interesting example of a Hegelian double negation. Overcoming the double nature of the adiabatic dark sector, at the end of the day we have a unified, non-adiabatic dark sector.}. In this way, looking for signatures of interaction provides a robust test of the $\Lambda$CDM standard model. The equation of state $p_{\Lambda} = - \rho_{\Lambda}$ for the dark energy component guarantees that it does not cluster, provided there is no momentum transfer between the components \cite{non-adiabatic,borges}. As we shall see, the latter happens in the case of a unified scalar field, and the above equation of state is a sufficient condition for dark energy to be strictly homogeneous.
In a recent paper \cite{sigma} we have shown that a free scalar field can indeed describe the dark sector in a unified way, playing the role of both dark energy and dark matter. Near the de Sitter limit, corresponding to the minimum of the scalar field potential, the solution found corresponds to a constant-rate energy flux from dark energy to dark matter. In this note we show that, around the de Sitter point, another particular solution exists, corresponding to a non-interacting fluid composed of conserved dark matter and a cosmological constant. We then also show that a general solution of the scalar field equations, valid for any time, is equivalent to a non-adiabatic generalised Chaplygin gas.

Let us consider a scalar field of mass $M$, with potential
\begin{equation}\label{4}
V(\phi) = V_{dS} + \frac{M^2 \phi^2}{2}.
\end{equation}
Its energy-momentum tensor is given by
\begin{equation}
T^{\mu\nu} = \partial^{\mu}\phi \, \partial^{\nu}\phi + \left( V - \frac{1}{2} \partial_{\alpha}\phi \, \partial^{\alpha}\phi \right) g^{\mu\nu}.
\end{equation}
We can formally decompose it into pressureless matter and a $\Lambda$ component, such that
\begin{eqnarray}\label{matter}
T_m^{\mu\nu} &=& \partial^{\mu}\phi \, \partial^{\nu}\phi, \\ \label{lambda}
T_{\Lambda}^{\mu\nu} = \Lambda  g^{\mu\nu} &=& \left( V - \frac{1}{2} \partial_{\alpha}\phi \, \partial^{\alpha}\phi \right) g^{\mu\nu}.
\end{eqnarray}
With this decomposition, the ``vacuum density" $\rho_{\Lambda}=\Lambda$ (which is not necessarily constant) is a covariant scalar, and the equation of state $p_{\Lambda}=-\rho_{\Lambda}$ remains the same for any observer. The covariant energy-momentum balance equations can be written as
\begin{eqnarray}\label{balance1}
T_{m;\nu}^{\mu\nu} &=& Q^{\mu},\\ \label{balance2}
T_{\Lambda;\nu}^{\mu\nu} &=& -Q^{\mu},
\end{eqnarray}
where
\begin{equation}
Q^{\mu} = Q u^{\mu} + \bar{Q}^{\mu},
\end{equation}
with
\begin{equation}
\bar{Q}^{\mu} u_{\mu} = 0.
\end{equation}
Here, $u^{\mu}$ is the fluid $4$-velocity, and $Q$ and $\bar{Q}^{\mu}$ represent, respectively, the energy and momentum transfer. Substituting (\ref{lambda}) into (\ref{balance2}), we have
\begin{eqnarray}\label{Q'}
Q &=& - \Lambda_{,\nu} u^{\nu},\\
\bar{Q}^{\mu} &=& \Lambda_{,\nu} (u^{\mu} u^{\nu} - g^{\mu\nu}).
\end{eqnarray}
A first order perturbation of these equations leads to $\delta \bar{Q}_0 = 0$ and, in a co-moving gauge, to
\begin{eqnarray}
\delta Q &=& -(\delta \Lambda)_{,0} \, ,\\
\delta \bar{Q}_i &=& (\delta \Lambda)_{,i}.
\end{eqnarray}
Therefore, if $\delta \bar{Q}^{\mu}$ is zero, $\Lambda$ is strictly homogeneous (and $\delta Q = 0$). Let us show that this is indeed the case.
From (\ref{lambda}) and (\ref{balance2}) we can write
\begin{equation}
Q_{\mu} = \partial_{\alpha}\phi \, \partial^{\alpha}\partial_{\mu}\phi - V' \,\partial_{\mu}\phi.
\end{equation}
In Fourier's space it takes the form
\begin{equation}
Q_{\mu} = ik_{\mu} (V' - k^2\phi) \phi,
\end{equation}
where $k_{\mu}$ is the scalar-field wavevector. In this derivation we have assumed an adiabatic regime, valid near the de Sitter point. In the non-adiabatic regime $Q_{\mu}$ presents additional terms due to the time dependence of $k_0$, but it is straightforward to show that it remains colinear to $k_{\mu}$. It was also assumed a space admitting plane waves as a complete ortogonal basis, which is the case of first-order perturbations of the spatially flat FLRW spacetime. By doing $i k_{\mu} = k u_{\mu}$ we can see that $\bar{Q}^{\mu} = 0$, that is, the matter component follows geodesics not only in the isotropic background but also in the perturbed space-time. With the potential (\ref{4}), we have\footnote{This can also be understood by taking the covariant derivative of equation (\ref{lambda}). Up to first order, it leads to
\begin{equation}
\Lambda_{,\nu} = \left[ m^2 - k^2 \left(1 + \frac{1}{2} \delta g_{00} \right) \right] k_{\nu} \phi^2,
\end{equation}
and hence to a homogeneous $\Lambda$ in a comoving frame.}
\begin{equation}\label{Q}
Q_{\mu} = k (M^2 - k^2) \phi^2 u_{\mu}.
\end{equation}

Let us turn now to the scalar-field solutions in the FLRW space-time. In the spatially flat case the Friedmann equations are given by
\begin{eqnarray}\label{7a}
3H^2 &=& V + 2H'^2,\\\label{7b}
\dot{\phi} &=& - 2 H',
\end{eqnarray}
where a prime means derivative w.r.t. the scalar field, and a dot means derivative w.r.t. the cosmological time. With the help of (\ref{matter}) and (\ref{lambda}), the Klein-Gordon equation
\begin{equation}\label{8}
\ddot{\phi} + 3H\dot{\phi} + V'= 0
\end{equation}
can be written as
\begin{equation}\label{9}
\dot{\rho}_m + 3 H \rho_m = - \dot{\Lambda} \equiv \Gamma \rho_m,
\end{equation}
where the last equality defines the rate of matter creation $\Gamma$. Using (\ref{7b}) and (\ref{8}) into (\ref{9}), we can obtain
\begin{equation}\label{10}
V'= (3H + \Gamma) H'.
\end{equation}
For a given potential, equations (\ref{7a}) and (\ref{10}) fully determine the solutions for $H(\phi)$ and $\Gamma(\phi)$.
Expanding them around the minimum of $V$, the second order solution for $H$ can be written as
\begin{equation} \label{11}
H = H_{dS} \, (1 + A \phi^2).
\end{equation}
Using (\ref{4}) and (\ref{11}) into (\ref{10}) we obtain, at the same order of approximation, a constant rate of matter creation,
\begin{equation}\label{12}
\Gamma = \frac{M^2}{2AH_{dS}} - 3H_{dS}.
\end{equation}
On the other hand, from (\ref{7b}) and (\ref{9}) we have
\begin{equation}
\Lambda'= 2\Gamma H',
\end{equation}
which, after integration, leads to
\begin{equation}\label{13'}
\Lambda = 2\Gamma H + \tilde{\Lambda},
\end{equation}
where $\tilde{\Lambda}$ is an arbitrary integration constant. As we shall see, by properly fixing $\tilde{\Lambda}$ we can either have a dark energy density scaling linearly with $H$ or a $\Lambda$CDM solution. 

Let us first take $\tilde{\Lambda} = 0$. This is the most natural choice, since there is no known energy scale fixing a different value. We then have, from (\ref{13'}),
\begin{equation}\label{13}
\Lambda = 2 \Gamma H \quad \quad \quad (\tilde{\Lambda} = 0).
\end{equation}
In the de Sitter limit $\phi \rightarrow 0$ it follows, from (\ref{4}), (\ref{7a}), (\ref{11}) and (\ref{13}), that
\begin{equation}\label{15}
V_{dS} = 3H_{dS}^2 = \frac{4\Gamma^2}{3}.
\end{equation}
Substituting (\ref{11}) into (\ref{7a}), retaining only terms up to second order in $\phi$ and using (\ref{15}) we get
\begin{equation} \label{16}
\frac{32\Gamma^2}{9} A^2 - \frac{8\Gamma^2}{3} A + \frac{M^2}{2} = 0.
\end{equation}
Furthermore, using (\ref{15}) into (\ref{12}) we obtain
\begin{equation}\label{17}
M^2 = 4 \Gamma^2 A.
\end{equation}
From (\ref{16}) and (\ref{17}) we then have $A = 3/16$, which, substituted back into (\ref{17}), leads to
\begin{equation}\label{19}
\Gamma^2 = \frac{4M^2}{3}.
\end{equation}
We see that $M$ completely determines, through (\ref{15}) and (\ref{19}), the matter creation rate $\Gamma$, the de Sitter horizon $H_{dS}$ and the potential (\ref{4}).
The cosmological solution with constant-rate creation of matter corresponds to an expanding space-time which tends asymptotically to a stable de Sitter universe \cite{humberto}. It was tested against precise observations at background and perturbation levels (that have also included the baryonic content, neglected here for mathematical convenience), showing a good approximation to the present universe \cite{PLB}. On the other hand, a vacuum term scaling linearly with $H$, as in (\ref{13}), is corroborated by QCD quantum field estimations in the expanding space-time \cite{sigma,Schutzhold}. Close enough to the de Sitter phase the matter density varies as $a^{-3/2}$, slower than conserved matter, due to the matter creation. This can be seen from (\ref{matter}), which in Fourier's space gives $\rho_m = k^2 \phi^2$. Near the de Sitter phase, the Klein-Gordon equation (\ref{8}) has the solution $\phi \propto \exp(-kt)$, with $k = 3H_{dS}/4$. Using the de Sitter scale factor $a \propto \exp(H_{dS}t)$, we obtain indeed $\rho_m \propto a^{-3/2}$. On the other hand, from (\ref{Q}), (\ref{15}) and (\ref{19}) it is easy to show that $Q = \Gamma \rho_m$, which is consistent with the definition of $\Gamma$ given by (\ref{9}) and (\ref{Q'}).

Let us consider now a second particular case, given by the choice $\Gamma = 0$ in (\ref{13'}), that is, no energy flux between the components, which implies a constant cosmological term $\Lambda = \tilde{\Lambda}$ and conserved dark matter. Taking the limit $\phi \rightarrow 0$ in (\ref{4}), (\ref{7a}) and (\ref{11}), we have
\begin{equation}\label{tilde}
V_{dS} = 3H_{dS}^2 = \tilde{\Lambda}.
\end{equation}
On the other hand, by doing $\Gamma = 0$ in (\ref{12}) we get $A=M^2/(2\tilde{\Lambda})$. Substituting into (\ref{11}), substituting the result into (\ref{7a}) and taking up to second order terms, we obtain an identity, provided that
\begin{equation}\label{tilde2}
\tilde{\Lambda} = \frac{4M^2}{3} \quad \quad \quad (\Gamma = 0),
\end{equation}
that is, $A = 3/8$. Once more, we can see that the mass of the scalar field completely determines the dynamics, through (\ref{tilde}) and (\ref{tilde2}). The solution of the Klein-Gordon equation (\ref{8}) is still given by $\phi \propto \exp(-kt)$, but now with $k=3H_{dS}/2$. Therefore, by using again the de Sitter scale factor $a \propto \exp(H_{dS} t)$, we obtain, from (\ref{matter}), $\rho_m \propto a^{-3}$, as expected for conserved matter. It is also easy to show from (\ref{Q}) that $Q = 0$, as should be in the case of a constant $\Lambda$. 

The particular solutions considered above, corresponding to the choices $\tilde{\Lambda} = 0$ and $\tilde{\Lambda} = 4M^2/3$, were shown to be valid close to the de Sitter asymptotic phase. For earlier times, we would expect that they are good approximations as well, expectation corroborated by the observational tests \cite{PLB}. In the first case, for early times the creation rate $\Gamma$ is subdominant as compared to the expansion rate $H$, and the matter density evolves in the usual way, scaling with $a^{-3}$ \cite{humberto,PLB}. In the second case, for times when the cosmological term departures significantly from a constant, it is subdominant as compared to the matter density, which would explain why the $\Lambda$CDM model presents a good fit to data. Both models are competitive and, despite some theoretical reasons supporting the former \cite{sigma,Schutzhold}, only observations would fix the integration constant $\tilde{\Lambda}$.

A more general solution for the system (\ref{7a}) and (\ref{10}), valid for any time, corresponds to the creation rate
\begin{equation} \label{gCg}
\Gamma = -\alpha \sigma H^{-(2\alpha+1)},
\end{equation}
where $\alpha > -1$ and
\begin{equation}\label{sigma}
\sigma = 3H_{dS}^{2(\alpha + 1)}
\end{equation}
are constants (in addition, we take $\alpha < 0$, in order to have a positive matter-creation rate). Substituting (\ref{gCg}) into the conservation equation (\ref{9}) and using the Friedmann equation $\Lambda + \rho_m = 3H^2$, it is easy to show that
\begin{equation} \label{Lambda}
\Lambda = \sigma H^{-2\alpha},
\end{equation}
and, furthermore, that
\begin{equation} \label{ChaplyginH}
H^2 = H_0^2 \left[\Omega_{\Lambda0} + \Omega_{m0} a^{-3(1+\alpha)}\right]^{\frac{1}{1+\alpha}}.
\end{equation}
Here, the subindex $0$ refers to the present time, and $\Omega_{m,\Lambda}$ are the relative densities of the components. The reader may identify in the above result the Hubble function of a generalised Chaplygin gas. It is a non-adiabatic gas, with zero sound speed, since we have shown that $\Lambda$, the only component with pressure, is not perturbed\footnote{It is a late-time non-adiabaticity. As there is no pressure perturbation, the non-adiabatic contribution is $\delta p_{nad} = -\omega \delta \rho$, where $\rho$ is the gas density, and $\omega$ is its equation-of-state parameter. For $a\ll1$ the gas behaves like incoherent matter, and $\delta p_{nad} \approx 0$.}. If we fix $\alpha = 0$ or $\alpha = -1/2$, we have, respectively, the $\Lambda$CDM model and the model with $\Lambda \propto H$, studied above.
Substituting (\ref{gCg}) into (\ref{10}) leads, after integration in $\phi$, to
\begin{equation}\label{V}
V = \frac{3H^2}{2} + \frac{\sigma}{2} H^{-2\alpha}.
\end{equation}
Substituting (\ref{V}) into (\ref{7a}) we obtain
\begin{equation}
4H'^2 -3H^2 + \sigma H^{-2\alpha} = 0.
\end{equation}
Substituting its solution back into (\ref{V}), we obtain $V(\phi)$. For $\alpha = 0$ (the $\Lambda$CDM model) we find\footnote{For $\alpha = -1/2$, the potential was found in \cite{borges}.}
\begin{equation}\label{exact}
V(\phi) = \frac{V_{dS}}{2} \left[ 1 + \cosh^2 \left( \frac{\sqrt{3}}{2} \phi \right) \right].
\end{equation}
For small $\phi$ we recover the potential (\ref{4}), with $M$ given by (\ref{tilde2}) and (\ref{tilde}).

We can show again that, once fixed the Chaplygin parameter $\alpha$, all the dynamics is determined by the mass of the scalar field. For that, it is enough to take the creation rate at the de Sitter point, given, from (\ref{gCg}) and (\ref{sigma}), by the constant
\begin{equation}\label{Gconst}
\Gamma = - 3 \alpha H_{dS} \quad \quad \quad (\textrm{de Sitter}).
\end{equation}
The cosmological term is given by (see (\ref{13'}))
\begin{equation}
\Lambda = V_{dS} = 2\Gamma H_{dS} + \tilde{\Lambda} \quad \quad (\textrm{de Sitter}).
\end{equation}
By doing $\Lambda = 3H_{dS}^2$, we obtain
\begin{equation}
\tilde{\Lambda} = 3H_{dS}^2 (1+2\alpha).
\end{equation}
If we now substitute our approximate solution (\ref{11}) into (\ref{7a}) and take only terms up to second order in $\phi$, we obtain
\begin{equation}\label{quadratica}
16H_{dS}^2 A^2 - 12H_{dS}^2 A + M^2 = 0.
\end{equation}
On the other hand, substituting (\ref{Gconst}) into (\ref{12}) we find
\begin{equation}\label{linear}
M^2 = 6 H_{dS}^2 (1-\alpha) A.
\end{equation}
Finally, from (\ref{quadratica}) and (\ref{linear}) we have $A = 3(1+\alpha)/8$, and
\begin{equation}\label{M}
M = \frac{3H_{dS}}{2}\sqrt{1-\alpha^2}.
\end{equation}
For a given $M$, near the de Sitter point the generalised Chaplygin gases with $-1 < \alpha < 1$ span the potentials of type (\ref{4}) with $V_{dS} \geq 4M^2/3$. For $\alpha = -1/2$ and $\alpha = 0$, we re-obtain the two particular cases discussed above. Near de Sitter we have, from (\ref{8}), $\phi  \propto \exp(-kt)$, with
\begin{equation}\label{k}
k = \frac{3H_{dS}}{2}(1+\alpha).
\end{equation}
From (\ref{Q}) we can check, using (\ref{M}), (\ref{k}) and (\ref{matter}), that $Q = \Gamma \rho_m$, with $\Gamma$ given by (\ref{Gconst}). 

The description of dark energy through a dynamical scalar field has been exhaustively studied, offering a theoretical alternative to the cosmological constant \cite{teoricos}. In this paper we have explored the possibility of both dark constituents, dark energy and dark matter, be modeled by a single scalar field. We have shown that the scalar field energy-momentum tensor can be split into a homogeneous component, responsible for the expansion acceleration, and a pressureless one that can be identified with clustering dark matter. In general, there is an energy flux between these components and, with a proper choice of the scalar field potential, this dark sector can be mapped into a non-adiabatic generalised Chaplygin gas, with the $\Lambda$CDM model as a particular case. In the de Sitter asymptotic limit, the matter creation rate is constant, and the scalar field has imaginary frequency, with wavelength of the order of the Hubble horizon. This last result implies, in particular, that the dark matter component has no well defined particle content and, as well as dark energy, is best interpreted as a field in the present context. This is natural, because we are describing both dark energy and dark matter in a unified way, as manifestations of the same fundamental field. It is also comprehensible on the light of the so called dark degeneracy, the theoretical arbitrariness in defining the dark components at the macroscopic level \cite{borges,kuns}. Although this degeneracy can be broken at the observational level by defining dark matter as the clustering component, as done in this paper, the dark sector constituents still have, in this realm, a common origin. This possibility is an interesting challenge for the present efforts made in the experimental detection of dark matter particles.

\newpage

\section*{Acknowledgements}

I am thankful to J. S. Alcaniz, H. A. Borges, J. C. Fabris, G. A. Mena Marug\'an and A. Saa for helpful comments. Work partially supported by CNPq, with grant no. 307467/2017-1.

\end{document}